# Radiation Transfer Models in Galaxies

Nikolaos D. Kylafis*† and Emmanuel M. Xilouris**

*University of Crete, Physics Department, P.O. Box 2208, 71003 Heraklion, Crete, Greece
†Foundation for Research and Technology-Hellas, P.O. Box 1527, 71110 Heraklion, Crete, Greece
**National Observatory of Athens, I. Metaxa & Vas. Pavlou str., Palaia Penteli, 15236, Athens, Greece

**Abstract.** The dust in galaxies makes radiation transport calculations in them absolutely necessary. It is not only common practice in Astrophysics, but also wisdom, to try to make as simple models as possible to simulate physical systems. For spiral galaxies, however, this turned out to be catastrophic. For years, the major question of the opacity of spiral galaxies kept the community divided, because the models were too simple. A spiral galaxy appears, to first order, to have exponential distributions of stars and dust, which cannot be approximated with uniform distributions. We will review the radiative transfer methods used in galaxies and we will comment on their pluses and minuses. We will also present some of the main results of the application of one of the methods to the observations.

## INTRODUCTION

We will start this review by asking the naive and rhetoric question of why radiation transfer models are necessary for the study of spiral galaxies. If galaxies were completely free of dust or if the existing dust had a negligible effect on the light emitted by the stars, then we could get by without radiation transfer models. However, as it is obvious in edge-on spiral galaxies, a significant fraction of the emitted light is absorbed by the dust. Furthermore, since the albedo of the dust grains is different than zero, light that would otherwise reach an observer is scattered out of the line of sight and light that is seen by an observer would not be seen if it weren't for the dust that scattered it into the line of sight. It is therefore obvious that radiation transfer models are necessary for the study of galaxies.

The next, not so naive, question is: Can we get by with simple models? It is customary for physicists and astronomers to try to explain the observed phenomena with as simple models as possible. In the case of spiral galaxies, however, astronomers have been misled by simple models.

One attempt was to study statistically a large number of galaxies using the empirical relation

$$\mu_{\text{obs}} = \mu_{\text{face-on}} - 2.5 C \log(a/b), \tag{1}$$

where $\mu$ stands for surface brightness (in mag/arcsec$^2$), $a/b$ is the axial ratio of the observed image of the galaxy and $C$ is a phenomenological parameter for the dust, varying from zero to one (see below) and is to be inferred from the observations. The above equation was claimed to relate the observed surface brightness $\mu_{\text{obs}}$ to the surface brightness $\mu_{\text{face-on}}$ that the galaxy would have if it were seen face-on.

In the one limit of $C = 1$, the above relation describes correctly the observed brightness of a *dustless* spiral galaxy with inclination. In the other limit of $C = 0$ it *supposedly* describes a completely opaque spiral galaxy, i.e., observed surface brightness independent of inclination. This, however, is completely wrong, because it makes the *unrealistic assumption that stars and dust are uniformly mixed* in the galactic disk. Then, in the case of an opaque spiral galaxy, the observer sees only the "skin" ($\sim 1$ mean free path) of the galaxy and measures the same surface brightness at all inclinations.

If dust were uniformly mixed with stars, then edge-on spiral galaxies would never exhibit dust lanes. Dust lanes appear in edge-on spiral galaxies because the scale height of the dust is significantly less than that of the stars.

Equation (1) has been used by many people, but a significant impact on the community had the work of Valentijn [35], who used a sample of $\sim 2600$ spiral galaxies of type $\sim$ Sb, and concluded that galaxy disks are opaque nearly throughout. Today we know that this is not correct, but the paper stimulated a lot of research. As Martin Rees has said, "it is more important to be stimulating than right".

Using the same database as Valentijn, but making more realistic assumptions about the dust distribution, Huizinga & van Albada [19] reached the opposite conclusion (see also [12]). They observed that the isophotal diameters of the galaxies increase with inclination and concluded that the face-on optical depth in the V band is significant only in the central regions of the galaxies.

Another stimulating paper was that of Disney, Davies, & Phillips [15]. Their basic proposal (slightly exaggerated here) was that galaxy disks could be completely opaque and we are fully unaware of it, because they behave like transparent systems. Imagine that the galactic plane is an opaque sheet of dust. Then in the face-on view we see half of its light, which is an acceptable value for a galaxy, and the isophotal diameters increase with inclination as they do in transparent galaxies.

Again, the observed dust lanes have convinced the community that Nature has decided not to use opaque sheets of dust in spiral galaxies. Nevertheless, the above paper and the meeting in Cardiff in 1994 (Davies & Burstein [11]) stimulated a lot of research on the subject of the opacity of spiral galaxies.

Many people (they will go unnamed here) have used the infamous "sandwich model" to draw conclusions about real galaxies. This is a classic case where the baby was thrown with the wash. Spiral galaxies are complicated structures, but to zeroth order they can be described by exponential distributions of stars and dust. These exponentials cannot be approximated with constants if the dynamic range is large, i.e., if the exponential extends over several scale lengths and scale heights.

During the meeting in Cardiff, a vote was taken among the $\sim 50$ participants. According to recollections, about half voted that galaxy disks are opaque and the rest that they are transparent, except possibly for their central regions. This led several people to the conclusion that detailed radiative transfer calculations in realistic galaxy models had to be done. Indeed, from then on, quite accurate and fairly realistic radiative transfer models of spiral galaxies started to be developed.

# RADIATION TRANSFER MODELS

Four methods have been used so far in radiation transfer models of spiral galaxies. We will present each one separately and in the end we will compare them to see their pluses and minuses.

## Method 1: Scattered intensities

The first attempt to do a somewhat realistic model of a spiral galaxy (NGC 891) was that of Kylafis & Bahcall [22]. The paper was largely ignored for several years, not because there was something fundamentally wrong with it, but because people thought that they could get by with simple models.

Systematic radiative transfer modeling of spiral galaxies using the method of scattered intensities was done by Byun et al. [8], Xilouris et al. ([41], [42], [43]), and Tuffs et al. [33].

To our knowledge, the method was first used by van de Hulst & de Jong [36] to explain the diffuse galactic radiation [37].

### *Description of the method*

In a medium that allows both scattering and absorption, one can write for the intensity $I$ seen by an observer along a given direction

$$I = I_0 + I_1 + I_2 + \cdots = I_0 + \omega I'_1 + \omega^2 I'_2 + \cdots, \qquad (2)$$

where $I_0$ is the unscattered intensity, $I_1$ is the once scattered intensity, $I_2$ is the twice scattered intensity, and $\omega$ is the albedo of the dust grains, which is shown explicitly in the second version. All of the above terms are wavelength dependent.

For the calculation of $I_0$ seen by an observer along a given direction, one needs to compute accurately a sum of the form

$$I_0 = \sum_{\text{all } \Delta s} \eta(s) \Delta s e^{-\tau(s)}, \qquad (3)$$

where $\eta(s)$ is the emissivity at position $s$ along the given direction, $\Delta s$ is a small interval along this direction, and $\tau(s)$ is the optical depth from position $s$ to the observer. Equation (3) says that only a fraction $e^{-\tau(s)}$ of the intensity $\eta(s)\Delta s$ produced in the interval $\Delta s$ escapes unscattered.

The calculation of $I_1$ seen by an observer along a given direction is a little more demanding, but still straightforward. In position $s$ along the line of sight $\hat{\mathbf{n}}$ consider the intensity $I_0(\hat{\mathbf{n}}')$ from all directions $\hat{\mathbf{n}}'$ that arrives there and let $p(\hat{\mathbf{n}}',\hat{\mathbf{n}})(\Delta\Omega/4\pi)$ be the probability that after one scattering the scattered light will go into solid angle $\Delta\Omega$ around $\hat{\mathbf{n}}$. Then, the intensity $I_1$ is given by a sum of the form

$$I_1 = \omega \sum_{\text{all } \Delta s} [\kappa \sum_{\text{all } \hat{\mathbf{n}}'} I_0(\hat{\mathbf{n}}') p(\hat{\mathbf{n}}',\hat{\mathbf{n}})(\Delta\Omega/4\pi) \Delta s] e^{-\tau(s)}, \qquad (4)$$

where the quantity in the square brackets times $\omega$ is the scattered intensity at position $s$ into the line of sight and $\kappa$ is the extinction coefficient. This way of writing of $I_1$ makes the analogy with $I_0$ complete.

Now the question arises: How important are the terms $I_n$ with $n \geq 2$? To answer it, let's make a worst case analysis. Let $I_0$, $I'_1$, $I'_2$, $\cdots$ (see Eq. 2) be of the same order of magnitude (say 1). Then, for $\omega = 0.5$, the entire series in Eq. (2) adds up to 2, while the $I_0 + \omega I'_1 = 1.5$. Thus, even a significant error in the computation of the terms $I_n$ with $n \geq 2$ implies a negligible (for Astrophysics) error in the total intensity $I$. Thus, the reasonable approximation

$$\frac{I_n}{I_{n-1}} \approx \frac{I_1}{I_0}, \quad n \geq 2, \tag{5}$$

allows one to compute the total intensity $I$ quite accurately with the use of $I_0$ and $I_1$ only, namely

$$I = \frac{I_0}{1 - I_1/I_0}. \tag{6}$$

## Method 2: Monte Carlo

This is by far the simplest method conceptually. It allows one to mimic exactly what Nature does. Several people have used this method in radiative transfer calculations in spiral galaxies ( [6], [13], [40], [4], [25]). The Monte Carlo method is quite old and it has been applied to various environments (e.g., [9], [38], [39], [21], [16]).

### *Description of the method*

The fundamental principle of the Monte Carlo method can be described as follows (Cashwell & Everett [9]):

Suppose that a process has 3 outcomes. Outcome A with probability 0.2, outcome B with probability 0.3, and outcome C with probability 0.5. If a large number $N$ of random numbers $r$ uniformly distributed in the interval [0,1) are produced, then we expect that approximately $0.2N$ will fall in the interval [0,0.2), $0.3N$ will fall in the interval [0.2, 0.5), and $0.5N$ will fall in the interval [0.5, 1). In other words, the value of a random number $r$ uniquely determines one of the three outcomes.

More generally, if $E_1$, $E_2$, $\cdots$, $E_n$ are $n$ independent, mutually exclusive events with probabilities $p_1$, $p_2$, $\cdots$, $p_n$ and $p_1 + p_2 + \cdots + p_n = 1$, then we will agree that a random number $r$ that satisfies the inequalities

$$p_1 + p_2 + \cdots + p_{i-1} \leq r < p_1 + p_2 + \cdots + p_i \tag{7}$$

determines event $E_i$.

For continuous distributions, the generalization goes as follows: Let $p(x)dx$ be the probability for event $x$ ($a \leq x < b$) to occur. Then, by picking a random number $r$ and setting

$$\int_a^x p(\xi)d\xi = r \tag{8}$$

one determines event *x* uniquely. The *x*'s thus selected obey the probability density $p(x)$.

*Procedure*

The procedure that one follows in a Monte Carlo code can then be described by the following steps:

Step 1: Consider a photon that was emitted at position $(x_0, y_0, z_0)$.

Step 2: To select a random direction $(\theta, \phi)$, pick a random number $r$ and set $\phi = 2\pi r$ and another random number $r$ and set $\cos\theta = 2r - 1$.

Step 3: To find the step $s$ that the photon makes before an event (scattering or absorption) occurs, pick a random number $r$ and the probability density

$$p(s) = (1/l)e^{-s/l}, \tag{9}$$

where $l$ is the mean free path of the medium, must be used in Eq. (8).

Step 4: The position of the event in space is at $(x, y, z)$, where

$$x = x_0 + s\sin\theta\cos\phi \tag{10}$$

$$y = y_0 + s\sin\theta\sin\phi \tag{11}$$

$$z = z_0 + s\cos\theta \tag{12}$$

Step 5: To determine what kind of event occurred, pick a random number $r$. If $r < \omega$, the event was a scattering and go to step 6, else it was an absorption and go to step 1.

Step 6: Determine the new direction $(\Theta, \Phi)$, where $\Theta$ is the angle between the old and the new direction (to be determined from the Henyey-Greenstein [18] phase function) and $\Phi = 2\pi r$.

Step 7: Convert $(\Theta, \Phi)$ into $(\theta, \phi)$ and go to step 3.

## Method 3: Solution of the radiative transfer equation by discretization

Let $\hat{\mathbf{n}}$ be the direction of interest and $s$ is the variable along it. The radiative transfer equation is

$$\frac{dI}{ds} = -\kappa(s)[I(s) - \omega \int_{(4\pi)} I(s, \hat{\mathbf{n}}')p(\hat{\mathbf{n}}', \hat{\mathbf{n}})(d\Omega'/4\pi)] + \eta(s), \tag{13}$$

where $\eta(s)$ is the emissivity of the stars and $\kappa(s)$ is the extinction coefficient. This equation is to be discretized and the resulting difference equations to be solved.

The method was used by Bruzual et al. [7] for a sandwich model and it was extended by Baes & Dejonghe [3] for a plane parallel geometry. It has never been applied to a realistic galaxy model.

## Method 4: Spherical harmonics

This is by far the fastest and the most elegant method. It expands the intensity in Legendre polynomials. The method was used by Di Bartolomeo et al. [14] for a sandwich model and it was extended to a plane parallel geometry by Baes & Dejonghe [3]. Unfortunately, it has never been used for a realistic model of a galaxy.

## Other methods

Other methods have been used for the study of circumstellar dust (see, e.g., [32] these proceedings). Unfortunately, the community of circumstellar disks and the community of spiral galaxies do not talk to each other.

## Comparison of the four methods

A detailed comparison was given by Baes & Dejonghe [3] and all methods give the same results within negligible errors. Only the Monte Carlo method and the scattered intensities one have been used to model real galaxies. The agreement between these two methods in realistic models of spiral galaxies is excellent [24].

Both methods are computer intensive. Which of the two is preferable, depends on the problem to be solved. Our opinion is that for statistical studies of *similar galaxies*, binned with say inclination angle interval $\Delta i = 5$ degrees, the Monte Carlo method is preferred. This is because with *one* Monte Carlo run, even if it is time consuming, one gets the 18 model images that are wanted.

For detailed modeling of *one* spiral galaxy, we would suggest the method of scattered intensities. This is because the observations allow the determination of the inclination angle *i* of the galaxy with an accuracy of $\pm 0.2$ degrees. Such fine binning with the Monte Carlo method requires a *huge* number of photons in order to have a statistically significant number of photons in all the pixels. The efficiency of the Monte Carlo method in such a case is $\sim 0.2/90 \sim 0.2\%$. Furthermore, due to their statistical nature, Monte Carlo images will have a large difficulty in converging to an observed galaxy image.

A second advantage of the method of scattered intensities is that it allows one to calculate the intensity exactly at the pixel positions. Thus, the comparison with the observations is straightforward.

## APPLYING THE METHOD OF SCATTERED INTENSITIES TO THE OBSERVATIONS

In order to use the method of scattered intensities to simulate spiral galaxies, one first has to determine the way that stars and dust are distributed within the galaxies. The simplest stellar emissivity (luminosity per unit volume) that one can use consists of an exponential (in both radial and vertical directions) disk and a bulge which produces an

$R^{1/4}$ law, namely

$$L(R,z) = L_s \exp\left(-\frac{R}{h_s} - \frac{|z|}{z_s}\right)$$
$$+ L_b \exp(-7.67 B^{1/4}) B^{-7/8}, \quad (14)$$

with $h_s$ and $z_s$ being the scalelength and scaleheight of the disk respectively and

$$B = \frac{\sqrt{R^2 + z^2(a/b)^2}}{R_e}, \quad (15)$$

with $R_e$ being the effective radius of the bulge and $a$ and $b$ the semi-major and semi-minor axis respectively. Here $L_s$ and $L_b$ are the normalization constants for the stellar emissivity of the disk and the bulge respectively. Modeling of the spiral structure was done by Misiriotis et al. [23].

For the extinction coefficient one can use a double exponential law, namely

$$\kappa_\lambda(R,z) = \kappa_\lambda^0 \exp\left(-\frac{R}{h_d} - \frac{|z|}{z_d}\right), \quad (16)$$

where $\kappa_\lambda^0$ is the extinction coefficient at wavelength $\lambda$ at the center of the disk and $h_d$ and $z_d$ are the scalelength and scaleheight respectively of the dust. The central optical depth of the model galaxy seen face-on is

$$\tau_\lambda^f = 2\kappa_\lambda^0 z_d. \quad (17)$$

## Results

In what follows, we will describe the results of Xilouris et al. ([41], [42], [43]), who modeled seven edge-on galaxies (UGC 2048, NGC 891, NGC 4013, IC 2531, UGC 1082, NGC 5529, and NGC 5907). Having defined the stellar and dust distributions as above, they performed the radiation transfer calculations using the method of scattered intensities (described earlier) and fitted the observations (the optical surface brightness of the above edge-on spiral galaxies). In the fitting procedure (a $\chi^2$ minimization), the radiation transfer was performed for a set of parameters and a model galaxy was formed. Then, the observed surface brightness was compared with the computed surface brightness from the model and a new set of parameters was found. This was repeated until a minimum in the $\chi^2$ value was reached.

A comparison between models and observed image is shown in Fig. 1. In the left panel, the I-band observations of all seven galaxies are presented. In the right panel, the model images of these galaxies in the I-band are shown with the same scale and sequence as in the left panel so that a direct comparison can be made between each pair. All the values of the parameters as well as a detailed description of the comparison between model and observations are given in [41], [42], [43]. A number of interesting conclusions have been drawn from the above work.

The authors have quantified the opacity of the galaxies in terms of their central face-on optical depth $\tau^f$. For all the galaxies they have found central face-on optical depths with values less than one in all the bands analyzed, indicating that the galaxies would appear to be transparent when seen face-on, despite their obvious dust lane in their almost edge-on orientation seen in the observations. The values of the optical depth in all bands are plotted in Fig. 2 for each galaxy. Different symbols indicate different bands. In particular, open circles correspond to the B-band, open squares to the V-band, open triangles to the I-band, solid squares to the J-band and solid triangles to the K-band.

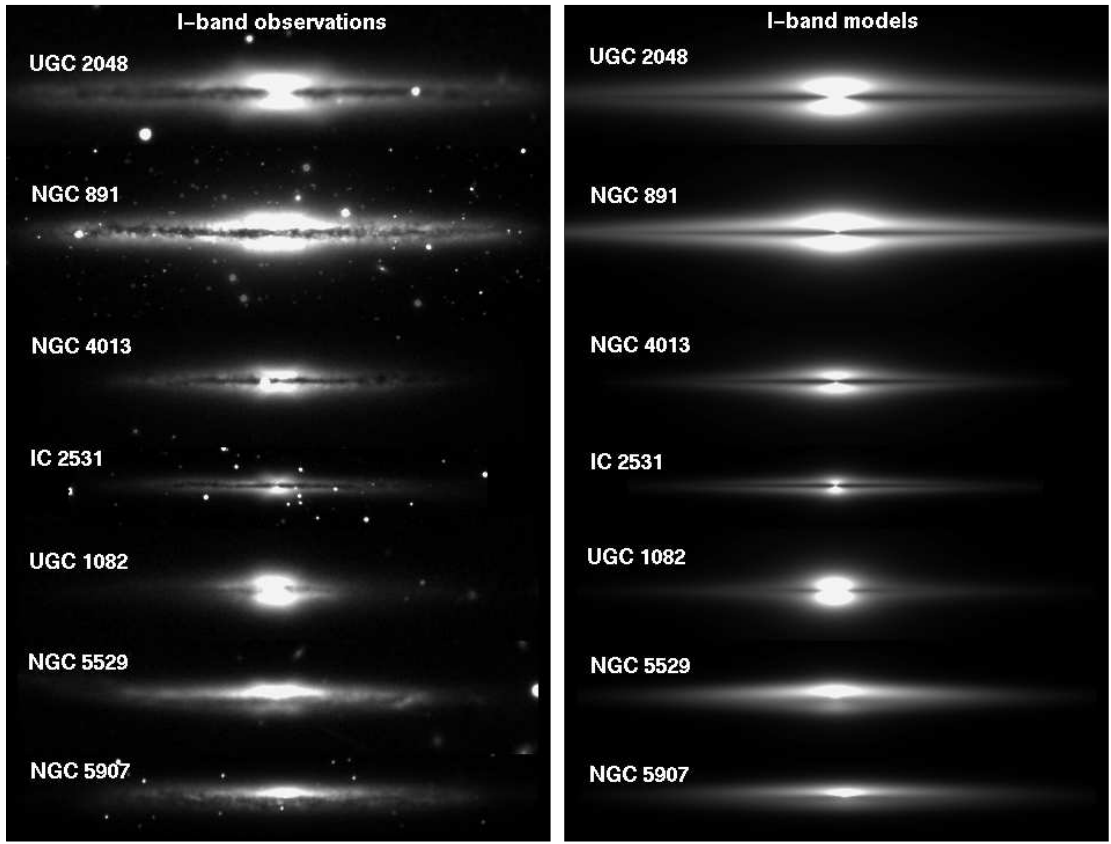

**FIGURE 1.** Left: I-band observations of the galaxies UGC 2048, NGC 891, NGC 4013, IC 2531, UGC 1082, NGC 5529, and NGC 5907 (top to bottom). Right: The respective I-band models for these galaxies.

It is of interest to see how the scaleheight of the stars is related to that of the dust. This is shown in Fig. 3, where the parameters $z_s$ and $z_d$ (in kpc) as well as their ratio are plotted for each galaxy. The symbols have the same meaning as in Fig. 2. Doing the statistics in the V-band data, the authors derive a mean ratio of $z_s/z_d = 1.8 \pm 0.6$ implying that the dust is more concentrated to the plane of the disk with respect to the stars. This was, of course, evident from the prominent dust lanes. This mean value of $z_s/z_d$ is plotted as a horizontal dashed line.

Of particular interest is the extended distribution of dust in the radial direction relative to the stars, as can be seen from the values that were derived for the scalelengths. In order to have a better view of how the scalelengths of the stars and the dust are related, a plot (Fig. 4) similar to that of the scaleheights described above, is given. The symbols here have exactly the same meaning as that in Fig. 2. The statistics that were done in the V-band, give a mean ratio of $h_d/h_s = 1.4 \pm 0.2$. This value is shown as a dashed horizontal line. All seven galaxies were found to have a dust distribution which is more extended in the radial direction than the distribution of the stars. This was a prediction of the model. The extended radial distribution of the dust has been confirmed for our Galaxy using COBE/DIRBE data [10]. Analysis of ISO data for several galaxies [1], [27], [28] as well as SCUBA observations of NGC 891 in $450\mu m$ and $850\mu m$ [2], also indicate a more extended radial distribution for the dust than for the stars.

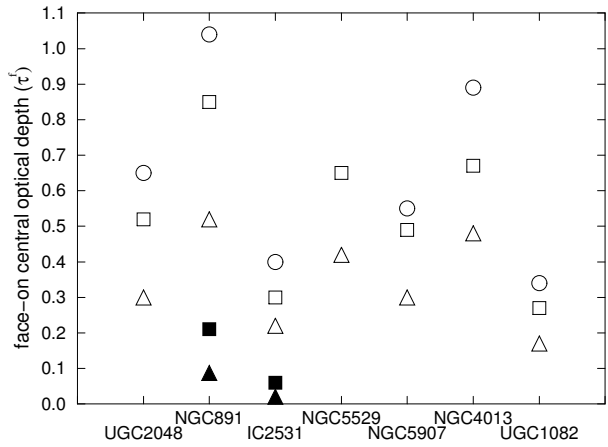

**FIGURE 2.** Values of the face-on central optical depth in all the bands that were modeled for each galaxy. Different symbols are for different bands. Open circles correspond to the B-band, open squares to the V-band, open triangles to the I-band, solid squares to the J-band, and solid triangles to the K-band.

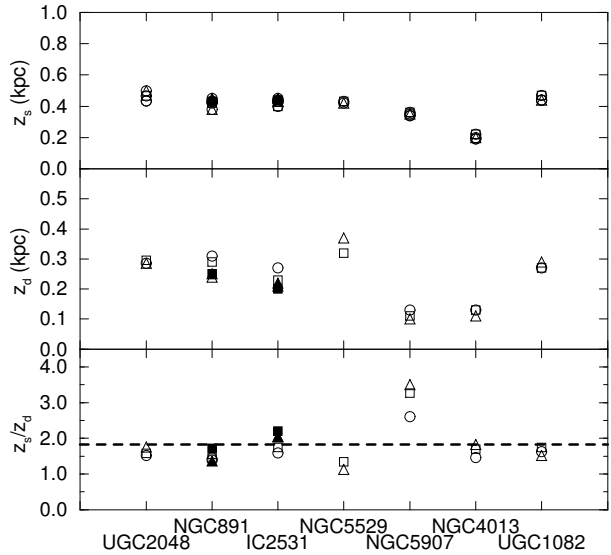

**FIGURE 3.** The values of $z_s$ (top graph), $z_d$ (middle graph), and $z_s/z_d$ (bottom graph) for each galaxy. All lengths are given in kpc. For the explanation of the symbols see Fig. 2. The dashed line in the last graph gives the mean value of the ratio calculated by using the V-band data.

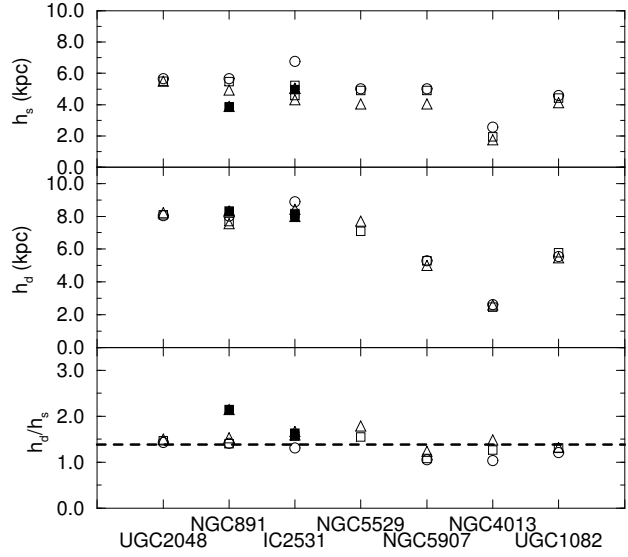

**FIGURE 4.** The values of $h_s$ (top graph), $h_d$ (middle graph), and $h_s/h_d$ (bottom graph) for each galaxy. All lengths are given in kpc. For the explanation of the symbols see Fig. 2. The dashed line in the last graph gives the mean value of the ratio calculated by using the V-band data.

Another interesting result that came out of their analysis was the dust mass. For all galaxies analyzed, the dust mass that was found gives a gas-to-dust mass ratio close to that derived for our Galaxy. More details are given in Fig. 5, where the gas mass (top graph) and the dust mass (middle graph), are plotted (in units of $M_\odot$) for each galaxy. For the dust mass (middle graph) the model calculations (filled circles) as well as the calculations using the IRAS data (open circles) are given. The gas-to-dust mass ratio (using the dust mass derived from the model) is also given for each galaxy (bottom graph). A mean value of this ratio is $M_g/M_d = 360 \pm 160$ (close to the value of $\sim 160$ adopted for our Galaxy, see [31], [30]) and is plotted as a horizontal dashed line. For all the galaxies the dust mass was found to be about an order of a magnitude more than that calculated using the IRAS fluxes, which proved the existence of a cold dust component that had gone undetected by IRAS. This is evident from the middle graph of Fig. 5, where the two dust masses (the one calculated from the model and the other calculated using the IRAS fluxes) are directly compared. The extinction model that the authors used, being independent of the dust temperature (both warm and cold dust contribute to the extinction of light), is able to calculate the diffuse dust contained within the galaxies which constitutes most of the dust contained in spiral galaxies. We caution however the reader that more dust may be hidden in the form of dust clouds and clumps as well as in the spiral arm configuration and may have gone undetected by the extinction model which only simulates the smooth dust distribution. If so, the mean ratio of the gas mass to the dust mass reported earlier provides an upper limit. Strong support to this argument (that vast amounts of cold dust exists in spiral galaxies) came with subsequent far-infrared observations at longer wavelengths, with the cold dust seen in emission ([1], [2], [5], [17], [20], [26], [34]).

Having calculated the values of $A_\lambda/A_V$ for all the galaxies, the values of the relative extinction in each band can be compared with the values derived for our Galaxy [29]. This is done in Fig. 6, where the values of the extinction ratios $A_\lambda/A_V$ are plotted as a function of inverse wavelength ($1/\lambda$). Each symbol here represents a different galaxy as shown inside the graph. One can see that the extinction law derived for all the galaxies is in very good agreement with that measured for our Galaxy indicating common dust properties.

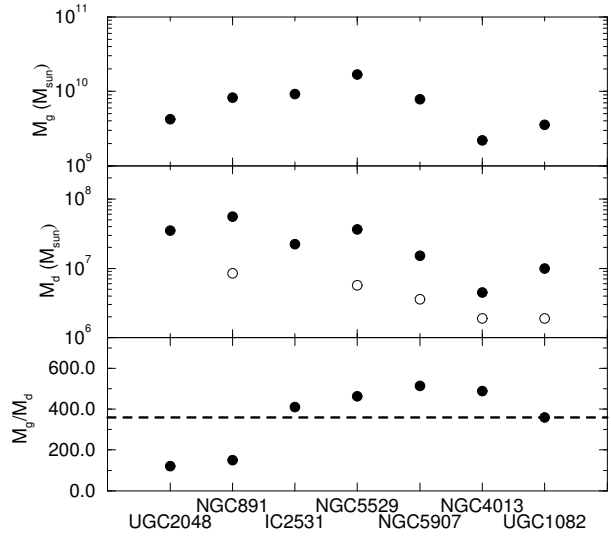

**FIGURE 5.** The values of $M_g$ (top graph), $M_d$ (middle graph), and $M_g/M_d$ (bottom graph) for each galaxy. All the masses are in units of $M_\odot$. For $M_d$ a direct comparison between the values as calculated by the model (filled circles) and those calculated using the IRAS fluxes (open circles), for whichever galaxy IRAS data were available, is given. The dashed line in the last graph gives the mean value of the gas-to-dust mass ratio.

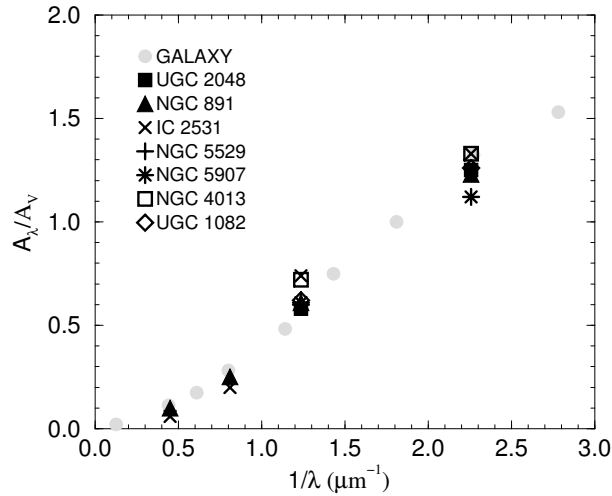

**FIGURE 6.** The observed (gray circles) values of $A_\lambda/A_V$ for our Galaxy and the values calculated for all the galaxies analyzed. The symbols are explained inside the plot.

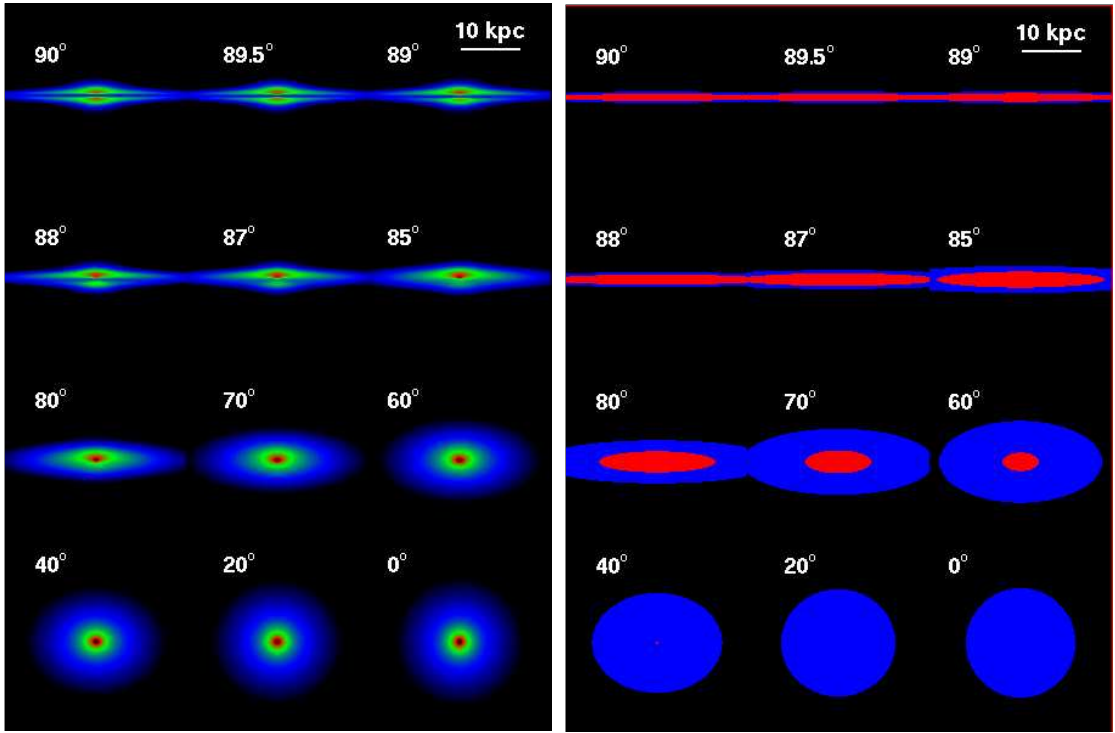

**FIGURE 7.** Left: A typical model galaxy in the B-band shown at various inclination angles (which are indicated on top of each model). Right: Mapping of the optical depth as a function of the inclination angle. White regions (red in electronic form) show the optically thick parts of the galaxy ($\tau > 1$), while grey regions (blue in electronic form) show the optically thin parts of the galaxy.

## A typical spiral galaxy

Assuming that the dust and the stellar disk follow an exponential 3D distribution and that the bulge is described by an $R^{1/4}$ law profile, the parameters (derived from the mean values) that describe a typical galaxy in the B-band are $z_s \approx 0.4$ kpc, $h_s \approx 5.0$ kpc, and $I_s \approx 20$ mag/arcsec$^2$ for the stellar disk and $z_d \approx 0.5 z_s$ and $h_d \approx 1.4 h_s$ for the dust disk. A mean B-band central face-on optical depth is $\tau^f \approx 0.8$. For the bulge, being more dependent on the morphological type of the galaxy, the values $a/b = 0.5$ for the ellipticity, $R_e = 1.5$ kpc for the effective radius, and $I_b \approx 12$ mag/arcsec$^2$ for the central edge-on surface brightness are adopted. Such a galaxy is shown in Fig. 7 (left panel) at various inclination angles, ranging from edge-on (90°) to face-on (0°). The faintest surface brightness level is 25 mag/arcsec$^2$. One thing that can clearly be seen is the existence of a dust lane down to almost 85°. The decrease of the optical diameter (going from edge-on to face-on) is also very obvious. This is the effect of integrating along shorter paths of light in the face-on configuration than in the edge-on one. In the right panel of Fig. 7 the "image" of the optical depth of such a galaxy is shown in different inclination angles. In each model-galaxy shown in this figure, the white regions (red in electonic form) show the parts of the galaxy that are optically thick ($\tau > 1$) and grey regions (blue in electronic form) show the optically thin parts of the galaxy. It is evident that the galaxy is optically thick, at least in the central regions, down to almost inclination angle of 60°, while the galaxy becomes totally transparent when viewed face-on.

## CONCLUSIONS

It has taken some time, but it has been realized by the community that *detailed* radiative transfer models of spiral galaxies are absolutely necessary. Several such models are now in use.